\documentclass{ws-ijmpd}
\usepackage[super,compress]{cite}
\begin{document}

\markboth{Authors' Names}
{Instructions for Typing Manuscripts (Paper's Title)}

%
\catchline{}{}{}{}{}
%

\title{A regular scale--dependent black hole solution}

\author{Ernesto Contreras\footnote{On leave from Universidad Central de Venezuela}}

\address{Departamento de F\'{\i}sica, Universidad de los Andes, Cra.1E
No.18A-10\\
Bogot\'a, Colombia\\
ej.contreras@uniandes.edu.co}

\author{\'Angel Rinc\'on}
 
\address{Instituto de F{\'i}sica, Pontificia Universidad Cat{\'o}lica de Chile,\\ Av. Vicu{\~n}a Mackenna 4860\\
Santiago, Chile.\\
}

\author{Benjamin Koch}
 
\address{Instituto de F{\'i}sica, Pontificia Universidad Cat{\'o}lica de Chile,\\ Av. Vicu{\~n}a Mackenna 4860\\
Santiago, Chile.\\}

\author{Pedro Bargue\~no}

\address{Departamento de F\'{\i}sica, Universidad de los Andes, Cra.1E
No.18A-10\\
Bogot\'a, Colombia\\
p.bargueno@uniandes.edu.co}

\maketitle

\begin{history}
\received{Day Month Year}
\revised{Day Month Year}
\end{history}

\begin{abstract}
In this work we present a regular black hole solution, in the context of scale--dependent General Relativity, satisfying the weak energy condition. 
The source of this solution is an anisotropic effective energy--momentum tensor which appears when the scale dependence of the theory is turned--on.
In this sense, the solution can be considered as a semiclassical extension of the Schwarzschild one.
\end{abstract}


\ccode{PACS numbers:}


\section{Introduction}\label{intro}

Black holes (BH) do exist in nature \cite{Abbott2016,Abbott2016bis}. However, their interior is plagued with uncertainties due to the presence of 
singularities which were predicted to occur time ago under some circumstances \cite{Hawking}. It is believed that, until a complete theory of quantum
gravity is developed, these singularities can be smoothed out by some effective theory operating, presumably, at the Planck scale. In this
singularity--free context, the interest in regular ({\it i. e.}, non--singular) BH solutions has considerably grown along the years. To our
knowledge, Sakharov \cite{Sakharov1966} and Gliner \cite{Gliner1966} were the first who conceived how to regularize a BH by introducing a non--singular 
de Sitter core inside it. After their pioneering work, the subsequent ideas presented by Bardeen \cite{Bardeen1968}, Dymnikova \cite{Dymnikova1992}, 
Ay\'on--Beato and Garc\'{\i}a \cite{Ayon1998} and Bronnikov \cite{Bronnikov2000,Bronnikov2001} have been closely followed in order to obtain regular BH solutions, 
many of them employing nonlinear electrodynamics as an appropriate matter source (see, for example, Ref. \cite{Sajadi2017} and references therein).
Regarding the plausibility of these regular BH solutions, the energy conditions play an important role. In this sense, Mars {\it et al.} 
\cite{Mars1996} were the first who address the posibility of finding a regular Schwarzschild--like solution satisfying the weak energy condition (WEC).
Recently, Balart and Vagenas \cite{vagenas2014,vagenas2015} have constructed some regular BH solutions satisfying the WEC using non--linear electrodynamics.

Currently, and in the context of quantum gravity, new ways to introduce quantum corrections to the well 
know BH solutions have been considered. One of these techniques assume improved solutions, i.e. after getting the classical solution one promotes 
the metric function to a scale dependent one, which is known from the renormalization
group flow (see \cite{Bonanno:2000ep,Bonanno:2006eu,Koch:2013owa,Koch:2014cqa} and references therein). Another different approach is based in 
the effective action. In particular, if we take into account an action where the coupling constants are not ``constant'' anymore, effective
Einstein field equations are obtained. Thus, the domain of the classical solution is extended by taking 
into account scale--dependent couplings. This technique has been considerably studied allowing us to recover the classical
solution at some limit and improving our knowledge about BHs. The idea was previously investigated by Weinberg through the well known Weinberg's Asymptotic
Safety program 
\cite{
Weinberg:1979,
Wetterich:1992yh,
Dou:1997fg,
Souma:1999at,
Reuter:2001ag,
Fischer:2006fz,
Percacci:2007sz,
Litim:2008tt}.
Recently, the technique has been used both in three--dimensional 
\cite{koch2016,Rincon:2017ypd,Rincon:2017goj,Rincon:2017ayr}
and four--dimensional \cite{koch2015b} BH physics (including a non--trivial scale--dependent polytropic black hole solution \cite{polytropic}).

In this work we study how a particular spherically symmetric and static regular BH solution obtained in Ref. \cite{vagenas2014} can be
interpreted in terms a scale--dependent gravitational theory \cite{koch2015,koch2015b,koch2016,Rincon:2017ypd,Rincon:2017goj} without invoking nonlinear
electrodynamics. In this sense, the solution here 
presented can be considered as a semiclassical ({\it i. e.}, scale--dependent) extension of the Schwarzschild one, providing an exact solution in the line 
of thought of Ref. \cite{Mars1996}.

The work is organized as follows. Section \ref{rbhs} introduces and summarizes a particular regular BH solution satisfying the WEC in
the context of nonlinear electrodynamics. Section \ref{sdg} is devoted
to briefly introduce the reader to the scale--dependent gravitational setting which is employed in Section \ref{rsdbhs} to obtain the
regular scale--dependent BH solution, whose main features are described along the section. Some concluding remarks are given in Section \ref{remarks}.

\section{A Particular regular black hole solution}\label{rbhs}
In this section we discuss some results reported in Ref. \cite{vagenas2014} in the context of regular BH 
solutions coupled to non--linear electrodynamics sources in spherically symmetric space--times. In this work
the line element is parametrized as 
\begin{eqnarray}\label{metricff}
ds^{2}&=&-f(r)dt^{2} + f(r)^{-1}dr^{2}+r^{2}d\Omega^{2},
\\
f(r)&=&1-\frac{2G_{0}m(r)}{r}
\end{eqnarray}
where $m(r)$ is the mass function. In order to construct regular BH metrics, $m(r)$ is expressed as
\begin{eqnarray}
m(r) = M_0 \frac{\sigma(r)}{\sigma_{\infty}}
\end{eqnarray}
where $G_{0}$ is Newton's coupling, $M_0$ is the classical mass, and the distribution function 
$\sigma(r)$ satisfies $\sigma(r)>0$ and $\sigma'(r)>0$ for $r\ge0$. Additionally,
$\sigma(r)/r\to0$ as $r\to0$ and $\sigma_{\infty}=\sigma(r\to\infty)$ is the normalization factor. The distribution is chosen in such a manner that $m(r)/r\to0$ when $r\to\infty$. 
An example of a distribution which leads to a regular solution is the log-logistic one reported in Ref. \cite{vagenas2014}, which is defined as
\begin{eqnarray}
\sigma(r)=M_{0}G_{0}\left(1+\frac{Q_{0}^2}{6M_{0}r}\right)^{-3}. 
\end{eqnarray}
Note that the $Q_{0}$ appearing in this distribution function can be interpreted as a classical charge only after an 
appropriate nonlinear electrodynamics is assumed \cite{vagenas2014}. Even more, this particular matter content fulfills the 
WEC, being asymptotically Reissner--Nordstr\"{o}m (however, this interpretation is not unique. In fact we will base our work 
in this idea). For this $\sigma(r)$, the lapse function can be written as
\begin{eqnarray}\label{metricsss}
f(r)=1-\frac{2G_{0}M_{0}}{r}\left(1+\frac{Q_{0}^2}{6M_{0}r}\right)^{-3}, 
\end{eqnarray}
its curvature invariants revealing that the solution is regular everywhere.
The particular choice of (5) is to be understood as one possible example, which is not necessarily
to be preferred over other choices for example those given in \cite{vagenas2014,vagenas2015}.

It is worth mentioning that when $Q_{0}\to0$, Eq. (\ref{metricsss}) corresponds 
to the Schwarzchild lapse function. In this sense, this solution can be considered as 
a correction of the Schwarzschild one. In this this work, we propose that such a correction arises 
as a consecuence of the scale--dependence of the parameters of the theory.

\section{Scale--dependent gravity}\label{sdg}
In the context of scale--dependent couplings, the effective Einstein--Hilbert action 
(taking into account a non-null cosmological coupling)  reads
\begin{eqnarray}\label{action}
\Gamma[g_{\mu\nu},k]=\int \mathrm{d}^{4}x\sqrt{-g}\left[\frac{1}{2\kappa_k}\left(R -2\Lambda_{k}\right)\right] +
 S_k,
\end{eqnarray}
where $G_{k}$ and $\Lambda_{k}$ stand for the scale--dependent gravitational and cosmological couplings,
respectively, whereas $\kappa_k \equiv 8 \pi G_k$ is the scale dependent Einstein coupling and $S_{k}$ is the action for the matter sector, which take into account the scale dependent effect encoded in the index $k$.
Please note that after any scale setting, the index $k$ becomes a scalar function of the coordinates.

Variations with respect to the metric $g_{\mu\nu}$ lead to the modified Einstein's field equations
\begin{eqnarray}\label{einstein}
G_{\mu\nu} +  \Lambda_{k} g_{\mu\nu} = \kappa_{k}(T^{eff})_{\mu\nu},
\end{eqnarray}
where
$(T^{eff})_{\mu\nu}$ is the effective energy momentum tensor defined as
\begin{eqnarray}\label{eff}
(T^{eff})_{\mu\nu} := (T_{\mu\nu})_{k} - \frac{1}{\kappa_{k}}\Delta t_{\mu\nu}.
\end{eqnarray}
In the above definition, $(T_{\mu\nu})_{k}$ corresponds to the matter energy--momentum tensor and depends on the energy scale $k$ and $\Delta t_{\mu\nu}$ is given by
\begin{eqnarray}\label{nme}
\Delta t_{\mu\nu}=G_{k}\left(g_{\mu\nu}\square -\nabla_{\mu}\nabla_{\nu}\right)G_{k}^{-1}.
\end{eqnarray}

Among various possibilities for the scale setting $k \rightarrow k(x)$,
the variational approach \cite{koch2015}
is particularly attractive, since it guarantees a minimal dependence on the arbitrary renormalization parameter $k$
\begin{align}\label{scale}
\frac{\mathrm{d}}{\mathrm{d}k}\Gamma[g_{\mu \nu}, k] =0.
\end{align}

In principle, if we combine Eq. \eqref{einstein} with the equation obtained from \eqref{scale}, 
we might obtain the fields involved. 
In particular the scale setting equation \eqref{scale} allows to determine the scalar function $k(x)$.
Inserting such a solution back into $\Gamma_k$, would, up to a boundary term, give the effective action which is in agreement
 with the given symmetries.
However, the lack reliable of knowledge of $\Gamma_k$ (e.g. the beta functions of quantum gravity) does frustrate such attempts.
This problem can be circumvented by promoting both $G$ and $\Lambda$ to field variables
and by imposing one additional constraint.

In this work we will  follow the approach employed in Refs. \cite{koch2015,koch2015b,koch2016}, where the ignorance on the scale dependence of the
coupling parameters was encoded by promoting $G_{k}$ and $\Lambda_{k}$ to independent fields, $G(x)$ and $\Lambda(x)$. 
Moreover, a static and spherically symmetric space--time, with a line element parametrized as
\begin{eqnarray}\label{metricp}
ds^2=-f(r)dt^2 + f(r)^{-1}dr^2+r^2 d\Omega^2 
\end{eqnarray}
is assumed.

Note that, replacing Eq. (\ref{metricp}) in Eq. (\ref{einstein}) we obtain two independent differential equations for the three independent fields
$f(r)$, $G(r)$ and $\Lambda(r)$. An alternative way to decrease the 
number of degrees of freedom consists in demanding some energy condition on $(T^{eff})_{\mu\nu}$. It is well known that
the null energy condition (NEC) is the less restrictive of the usual energy conditions and that it can help us
to obtain suitable solutions of Einstein's field equations \cite{koch2016}. For the effective energy--momentum tensor, 
$(T^{eff})_{\mu\nu}$, the NEC reads
\begin{eqnarray}
(T^{eff})_{\mu\nu} \ell^{\mu}\ell^{\nu}= \left[T_{\mu\nu} - \frac{1}{\kappa(r)}\Delta t_{\mu\nu}\right]\ell^{\mu}\ell^{\nu} \ge 0,
\end{eqnarray}
where $\ell^{\mu}$ is a null vector. Considering the special case $\ell^{\mu}=\{ f^{-1/2}, f^{1/2} , 0, 0 \}$,
we obtain that $T_{\mu\nu}\ell^{\mu}\ell^{\nu}=0$ which implies that $\Delta t_{\mu\nu}\ell^{\mu}\ell^{\nu}\ge 0$.
However, it can be shown that $G_{\mu\nu}\ell^{\mu}\ell^{\nu}=0$ and, for consistency with Eq. (\ref{einstein}), we
demand
\begin{eqnarray}\label{necnm}
\Delta t_{\mu\nu}\ell^{\mu}\ell^{\nu}=0.
\end{eqnarray}
Note that Eq. \eqref{necnm} can be used to obtain an ordinary differential equation to solve for the gravitational coupling, $G(r)$, as was
previously indicated in Ref. \cite{Rincon:2017ayr}. The corresponding equation is given by
\begin{align}\label{EDO_G}
2\left[\frac{\mathrm{d} G(r)}{\mathrm{d} r}\right]^2 = G(r)\frac{\mathrm{d}^2 G(r) }{\mathrm{d} r^2 }.
\end{align}
Thus, solving \eqref{EDO_G} we get

\begin{eqnarray}\label{gr}
G(r)=\frac{G_{0}}{1+\epsilon r},
\end{eqnarray}
where $\epsilon\ge 0$ is a parameter with dimensions of inverse of length. 

One notes that for $r\rightarrow \infty $ one finds that $G(r) \rightarrow0$, which might be naively interpreted as an 
asymptotically free theory.
There is nothing particularly bad about asymptotically free theories or solutions.
However, one should be more careful with this interpretation, because $G(r)$ alone is not directly observable, e.g. by
an experiment using geodesics. 
Such experiments are only sensible to the line element $f(r)$, which would infer an effective Newton coupling,
which is not necessarily identical to the $G(r)$ in equation \eqref{gr}.

By using Eq. \eqref{nme} and Eq. \eqref{EDO_G}, the tensor $\Delta t_{\mu \nu}$ 
can be written such as $\Delta t^{0}_{0} = \Delta t^{1}_{1}$ and $\Delta t^{2}_{2} = \Delta t^{3}_{3}$ and explicitly we obtain
\begin{align} 
\Delta t^{1}_{1} &= -\frac{1}{2r}f(r) \left[4 + r \frac{\mathrm{d}}{\mathrm{d}r}\ln(f(r))\right]\frac{\mathrm{d}}{\mathrm{d}r}\ln(G(r)),
\label{Delta_tUd_gen_1}
\\
\Delta t^{3}_{3} &= - \hspace{0.08cm}\frac{1}{r} \hspace{0.08cm} f(r) \left[1 + r \frac{\mathrm{d}}{\mathrm{d}r}\ln(f(r))\right]\frac{\mathrm{d}}{\mathrm{d}r}\ln(G(r)).
\label{Delta_tUd_gen_2}
\end{align}
Thus, one note that the classical Einstein's field equations are recovered
when the logarithmic derivative of the gravitational coupling is constant according to Eqs. \eqref{Delta_tUd_gen_1} and \eqref{Delta_tUd_gen_2} , namely, in the limit 
$\epsilon\to 0$, $G(r)=G_{0}$, which implies $\Delta t_{\mu\nu}=0$.

In this sense, 
when $\epsilon$ vanishes, the running of the coupling parameters is turned--off. For this reason, $\epsilon$ is called the running parameter
\cite{koch2016,Rincon:2017ypd,Rincon:2017goj,Rincon:2017ayr}.
For any lapse function, in particular one with a regular behavior at the origin,
 the equations \eqref{einstein} can be solved
for the corresponding ``matter'' stress energy tensor
which satisfies $T^0_0 = T^1_1$ and $T^2_2 = T^3_3$, therefore
\begin{align}
\kappa(r)T^1_1 &= \frac{1}{r^2}f\left[1 + r \frac{f'}{f}\right]
- 
\frac{1}{2r}f\left[4 + r\frac{f'}{f}\right]\frac{G'}{G},
\\
\kappa(r)T^3_3 &= \frac{1}{r}f' \left[1+\frac{1}{2}r\frac{f''}{f'}\right]
-
\frac{1}{r}f\left[1+r \frac{f'}{f}\right]\frac{G'}{G},
\end{align}
with this in mind, we will construct an energy--momentum tensor which take into account the scale dependent effect through the gravitational coupling $G(r)$.

\section{Regular scale--dependent black hole solution}\label{rsdbhs}

\begin{figure}[ht!]
\centering
\includegraphics[width=\linewidth]{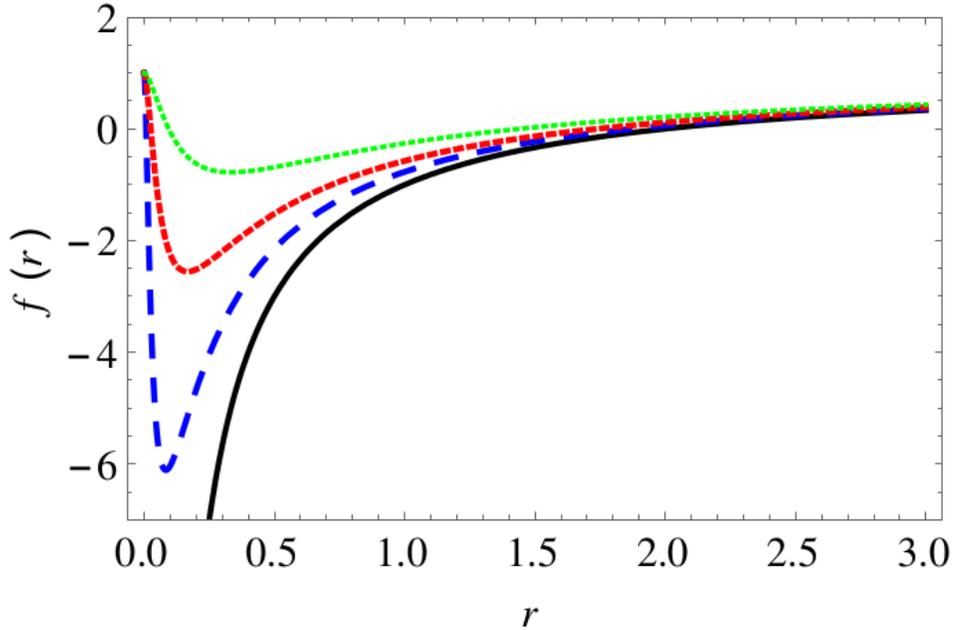}
\caption{\label{lapsefig}
Lapse function for
$\epsilon=0.00$ (black solid line), $\epsilon=0.25$ (dashed blue line), $\epsilon=0.50$
(short dashed red line) and $\epsilon=1.00$ (dotted green line). For illustrative purposes, $M_{0}$ and $G_{0}$ have been taken as unity.
}
\end{figure}

In this section we will construct a scale--dependent lapse function from Eq. (\ref{metricsss}). As commented at the end of section \ref{rbhs},
given the fact that when $Q_{0}\rightarrow0$ the Schwarzschild solution is recovered, we propose to reinterpret this correction arising as a consequence
of the scale--dependence of the theory. In this sense, $Q_{0}$ plays the role of the running parameter, $\epsilon$. Once the dimensions have been
correctly taken into account, the appropriate replacement is
\begin{eqnarray}\label{reemplazo} 
\frac{Q_{0}^2}{M_{0}}&\rightarrow& M_{0}^{2}G_{0}^{2}\epsilon.
\end{eqnarray}
Therefore, by the substitution given by Eq. (\ref{reemplazo}), the lapse function of Eq. (\ref{metricsss}) reads
\begin{eqnarray}
f(r)=1-\frac{2 G_{0} M_{0}}{r}\left(1+\frac{G_{0}^2 M_{0}^2 \epsilon }{6 r}\right)^{-3},
\end{eqnarray}
whose profile is shown in Fig. \ref{lapsefig}.

At this point, a number of comments are in order. First, the scale dependent solution 
has a de Sitter behaviour for $r\to0$, given by
\begin{eqnarray}\label{fdr}
f(r)=1-\frac{432 r^2}{G_{0}^5 M^5 \epsilon ^3}+\mathcal{O}(r^3), 
\end{eqnarray}
in complete agreement with the conditions demanded in Ref. \cite{dymnikova2003} to obtain regular solutions with a de Sitter center. From Eq. (\ref{fdr}) is easy
to see that an effective cosmological constant, given by $\Lambda_{eff}=\frac{1296}{G_{0}^5 M_{0}^5\epsilon^3}$ appears at $r\rightarrow 0$. Second, the running
solution is asymptotically Schwarzschild (see Fig. \ref{lapsefig}). And third, depending on the values for ($M_{0},\epsilon$), the solution presents
one (extremal case) or two (Cauchy and event) horizons. Even more, for certain combinations of ($M_{0},\epsilon$), no horizon appears.

A long but straightforward calculation shows that the scalar curvature, Ricci squared
and Kretschmann invariants are given in terms of $\epsilon$ as
\begin{eqnarray}
R&=&\frac{5184 G_{0}^5 M_{0}^5 \epsilon ^2}{\left(G_{0}^2 M_{0}^2 \epsilon +6 r\right)^5}\\
Ricci^2&=&\frac{6718464 G_{0}^6 M_{0}^6 \epsilon ^2 \left(G_{0}^4 M_{0}^4 \epsilon ^2+36 r^2\right)}{\left(G_{0}^2 M_{0}^2 \epsilon +6 r\right)^{10}}\\
\mathcal{K}&=&\frac{4478976 G_{0}^2 M_{0}^2}{\left(G_{0}^2 M_{0}^2 \epsilon +6 r\right)^{10}} \bigg(G_{0}^8 M_{0}^8 \epsilon ^4
+126 G_{0}^4 M_{0}^4 r^2 \epsilon ^2\nonumber\\
&&-216 G_{0}^2 M_{0}^2 r^3 \epsilon +648 r^4\bigg)
,
\end{eqnarray}
showing that the solution is regular everywhere. 
Note that, in contrast to 
the Schwarzchild case, both the  Ricci and the Ricci squared scalars are non--vanishing quantities.
Moreover, an expansion for small $\epsilon$ results in a deviation of order $\epsilon^2$ with respect to the
classical values. Regarding the Kretschmann scalar, an expansion for small values in the running
parameter leads to a modification of order $\mathcal{O}(\epsilon)$ with 
respect to the non--running case. It is clear that the so--called classical results are obtained in the limit $\epsilon\to0$, as expected.

Once a regular Schwarzschild--like solution has been obtained, now the key point is to reinterpret it as a solution of the scale--dependent Einstein's field 
equations given by Eq. (\ref{einstein}) without cosmological term. Given the intriguing appearance of the effective cosmological constant, 
one possibility is to associate the matter content with certain anisotropic {\it vacuum} 
which appears in a regime where the scale dependence cannot be ignored. In this sense, this {\it vacuum} should vanishes in the classical case, {\it i. e.}
for $\epsilon\rightarrow0$. This can be shown by an explicit caculation which leads to
\begin{eqnarray}\label{matterrun}
T^{0}_{0}&=&T^{1}_{1}=\bigg(G_{0}^8 M_{0}^8 \epsilon ^4+24 G_{0}^6 M_{0}^6 r \epsilon ^3+216 G_{0}^4 M_{0}^4 r^2 \epsilon ^2\nonumber\\
&&-648 r \left(G_{0}^3 M_{0}^3+3 G_{0} M_{0} r^2-2 r^3\right)\nonumber \\
&&+432 G_{0}^2 M_{0}^2 r^2 \epsilon  (2 r-3 G_{0} M_{0})\bigg)\frac{\epsilon  }{4 \pi  G_{0} r 
\left(G_{0}^2 M_{0}^2 \epsilon +6 r\right)^4},
\\
T^{2}_{2}&=&T^{3}_{3}= \bigg(G_{0}^{10} M_{0}^{10} \epsilon ^5+30 G_{0}^8 M_{0}^8 r \epsilon ^4+360 G_{0}^6 M_{0}^6 r^2 \epsilon ^3\nonumber\\
&&+432 G_{0}^4 M_{0}^4 r^2 \epsilon ^2 (5 r-6 G_{0} M_{0})
+7776 r^2 \left(G_{0}^3 M_{0}^3+r^3\right)\nonumber\\
&&-1296 G_{0}^2 M_{0}^2 r \epsilon  
\left(G_{0}^3 M_{0}^3-5 r^3\right)\bigg)\frac{\epsilon }{8 \pi  G_{0} r \left(G_{0}^2 M_{0}^2 \epsilon +6 r\right)^5}.
\end{eqnarray}

We note that this $T_{\mu\nu}$ has a divergence at $r\to0$ which is in apparent disagreement with one of the requirements demanded
in Ref. \cite{dymnikova2003} to have regular solutions with a de Sitter center. However, the non--matter energy momentum tensor, $\Delta t_{\mu\nu}$,
given by 
\begin{eqnarray}
\Delta t^{0}_{0}&=&\Delta t^{1}_{1}= \bigg(G_{0}^8 M_{0}^8 \epsilon ^4+24 G_{0}^6 M_{0}^6 r \epsilon ^3\nonumber\\
&&+216 G_{0}^4 M_{0}^4 r^2 \epsilon ^2
+216 G_{0}^2 M_{0}^2 r^2 \epsilon  (4 r-3 G_{0} M_{0})\nonumber\\
&&-648 r^3 (3 G_{0} M_{0}
-2 r)\bigg)\frac{2 \epsilon }{r (r \epsilon +1) \left(G_{0}^2 M_{0}^2 \epsilon +6 r\right)^4}\\
\Delta t^{2}_{2}&=&\Delta t^{3}_{3}=\bigg(G_{0}^8 M_{0}^8 \epsilon ^4+24 G_{0}^6 M_{0}^6 r \epsilon ^3
+216 G_{0}^3 M_{0}^3 r^2 \epsilon  (G_{0} M_{0} \epsilon -6)\nonumber\\
&&+864 G_{0}^2 M_{0}^2 r^3 \epsilon +1296 r^4\bigg)\frac{\epsilon  }{r (r \epsilon 
+1) \left(G_{0}^2 M_{0}^2 \epsilon +6 r\right)^4},
\end{eqnarray}
acts as a counterterm which leads to a regular effective energy momentum tensor whose components are expressed as
\begin{eqnarray}
(T^{eff})^{0}_{0}=(T^{eff})^{1}_{1}&=&-\frac{162 G_{0}^2 M_{0}^3 \epsilon  (r \epsilon +1)}{\pi  \left(G_{0}^2 M_{0}^2 \epsilon +6 r\right)^4}\\
(T^{eff})^{2}_{2}=(T^{eff})^{3}_{3}&=&-\frac{162 G_{0}^2 M_{0}^3 \epsilon  (r \epsilon +1) }{\pi  
\left(G_{0}^2 M_{0}^2 \epsilon +6 r\right)^5} \nonumber\\
&&\times\left(G_{0}^2 M_{0}^2 \epsilon -6 r\right).
\end{eqnarray}

Note that the effective density $-(T^{eff})^{0}_{0}$ is regular everywhere and decays faster that $r^{-3}$, as required in Ref. \cite{dymnikova2003}.
Moreover, in the limit $r\rightarrow 0$, this $T^{eff}_{\mu\nu}$ becomes isotropic and corresponds to the usual vacuum energy associated to the effective 
cosmological constant given by $\Lambda_{eff}=\frac{1296}{G_{0}^5 M_{0}^5\epsilon^3}$, as commented before. 
Other way of reobtaining the classical solution is taking the limit $\epsilon\to0$ to observe that the effective energy--momentum tensor vanishes 
according to Birkhoff's theorem.

Note that the effective energy--momentum tensor is composed of both matter and non--matter sectors. Although it is not 
mandatory that the non--matter sector fulfills any energy condition, we point out that the matter sector here considered
violates the WEC for large $r$ as can be shown in 
Eq. \ref{matterrun} (the $r^3$ term is the responsible for the violation of the WEC). In this sense, this matter content might be 
considered to be of quantum nature. As stated in
\cite{Hawking}, the dominant energy condition is violated due to the regularity of the solution.

\section{Concluding remarks}\label{remarks}

In this paper we have obtained a regular black hole solution, in the context of scale--dependent gravity, inspired by previous regular solutions based
on non--linear electrodynamics. Although the metric is formally the same in both situations, we would like to remark that, from the physical point of
view, their interpretation differs. In our case, the effective matter content, which we phenomenologically interpret as a kind of vacuum energy, 
modifies the well known Schwarzschild geometry, as a consequence of the scaling of the theory. In this sense, our interpretation is closer to
other approach based also on effective equations obtained from a variational principle \cite{Ziprick2010}. Specifically, the resemblance between our
work and that reported on Ref. \cite{Ziprick2010} is the appearance of an $r$--dependent Newton's constant and an effective energy--momentum
tensor, which is taken as nonvanishing when the mass function is non--constant \cite{Ziprick2010}, in contrast to our case where it is associated to the running 
parameter. However, although in both cases the obtained solution corresponds to a static and spherically symmetric regular black hole satisfying
the weak energy condition, the existence of
a counterterm which gives place to a regularized effective energy--momentum tensor is an intrinsic feature of the approach we have 
followed. As a final comment, we would like to point out that, given any spherically symmetric line element parametrized 
Eq. (\ref{metricp}), the whole procedure can be repeated. Within this parametrization, the key point is that the null
energy condition gives place to 
$G(r)$ given by Eq (\ref{gr}) and, therefore, the problem turns into an algebraic one for the components of the matter 
energy--momentum tensor. Therefore, the particular choice for the metric employed in this work is only a matter of simplicity. 
We could have chosen any other metric parametrized as previously commented. For example, in the Hayward black hole case 
\cite{Hayward2006} there is a parameter which controls the length scale below which quantum effects dominate. In this case, 
our running parameter will correspond to this scale and our procedure will provide a scale--dependent mechanism for the Hayward 
spacetime.
\section*{ACKNOWLEDGEMENTS}

The author A. R. was supported by the  CONICYT-PCHA/Doctorado  Nacional/2015-21151658. The author B. K. was supported by the 
Fondecyt 1161150. The author P. B. was supported by the Faculty of Science and Vicerrector\'{\i}a de Investigaciones of Universidad de los Andes, Bogot\'a, Colombia. P. B. dedicates this work to In\'es Bargue\~no--Dorta. The authors would like to ackowledge the
referees for very useful suggestions and comments which have serve to improve the quality of the manuscript.

\end{document}